\documentclass[twocolumn,amssymb,aps,prl,longbibliography]{revtex4-2}
\usepackage{amsfonts}
\usepackage[utf8]{inputenc}
\usepackage[english]{babel}
\usepackage{amsmath}
\usepackage{graphicx}
\usepackage{color}
\usepackage{amsmath}
\usepackage{bm}
\usepackage{xcolor}
\usepackage{array}
\usepackage{multirow}
\usepackage[normalem]{ulem}

\usepackage[colorlinks=false, linkbordercolor=red, citebordercolor=green, urlbordercolor=cyan, pdfborderstyle={/S/U/W 1}]{hyperref}

\begin{document}

\title{Coherent enhancement of collection of light from linear ion crystals}
\author{T.~D.~Tran$^1$, D.~Babjak$^1$, A.~Kovalenko$^1$, K.~Singh$^1$, M.~T.~Pham$^2$, P.~Obšil$^1$, A.~Lešundák$^1$, O.~Číp$^2$, L.~Slodička$^1$}
\address{$^1$ Department of Optics, Palacký University, 17. listopadu 12, 771 46 Olomouc, Czech Republic\\
$^2$ Institute of Scientific Instruments of the Czech Academy of Sciences, Královopolská 147, 612 64 Brno, Czech Republic}

\begin{abstract}
The efficient detection of light from trapped ions in free space is paramount for most of their applications. We propose a scheme to enhance the photon collection from linear ion strings. It employs the constructive interference of light scattered from ions along the axial direction in linear Paul traps. The coherent enhancement of photon collection is numerically optimized for a range of feasible spatial angles and realistic ion positions in a single harmonic Coulomb potential. Despite the large mutual distance of scatterers on the order of many wavelengths of scattered light, presented experimental tests confirm the feasibility of enhancements by a factor of $3.05 \pm 0.09$ with a crystal of nine $^{40}$Ca$^+$~ions. Further significant improvements using different ion species, which allow for suppression of the sensitivity to the residual thermal motion, are predicted. The proposed collection geometry is intrinsic to diverse linear ion trap designs and the methodology can be directly applied to an observation of scattering from ion crystals prepared in collective electronic excitations.
\end{abstract}


\maketitle

\section{Introduction}

Realizing an efficient interface between light and atoms represents one of the most active research directions in experimental quantum optics. While collective interactions between light and atoms can provide a feasible solution in large atomic ensembles~\cite{guerin2017light,sangouard2011quantum,chaneliere2018quantum}, small trapped ensembles of individual atoms or ions typically utilize a complementary approach based on high numerical aperture~(NA) collection optics. Diverse optical and trapping designs for efficient coverage of large solid angles with bulk optics~\cite{tey2009interfacing,alt2002objective,nelson2007imaging,gerber2009quantum,bakr2009quantum,gross2021quantum,wong2016high,fuhrmanek2011free,shu2011efficient,maiwald2012collecting,zarraoa2024quantum,araneda2020panopticon} or trap-embedded optical micro-devices~\cite{heine2009integrated,chou2017note,clark2014characterization,streed2012absorption} have been developed and tested. The implementations of such couplings have been typically significantly more challenging for trapped ions due to the high sensitivity of their motion to the proximity of dielectric surfaces, which limits the acceptable working distances of collection optics. At the same time, however, the tight and extremely stable trapping potentials of Paul traps make this platform superior in diverse fundamental tests and a broad range of applications~\cite{leibfried2003quantum,duan2010colloquium,ludlow2015optical,chou2017preparation,bruzewicz2019trapped,monroe2021programmable,postler2022demonstration,krutyanskiy2023telecom}. The availability of an efficient interface between internal electronic states and well defined photonic mode plays a critical role in a majority of their implementations. Recent progress of interfaces based on optical cavities demonstrated a possible viable solution~\cite{takahashi2020strong,casabone2015enhanced,krutyanskiy2023telecom}, however, the selection of a particular spatial mode structure and slow cavity decay rates impose limits on applications which rely on inherently fast dynamics~\cite{bushev2013shot,cerchiari2021measuring}. On the other hand, setups employing high-NA collection optics in free space are mostly limited to overall collection efficiencies on the order of~1-10~\%, even in particularly optimized trapping and imaging geometries~\cite{maiwald2012collecting,araneda2020panopticon,chou2017note,clark2014characterization,streed2012absorption}. These often result in severe restrictions of other critical implementation parameters, including the trapping potential structure, motional heating rates, or restricted field of view corresponding to only single or few trapped ions.

Here, we devise and demonstrate a scheme for enhancing the collection efficiency of light from linear ion strings by tailoring their far-field scattering patterns in the limit of small collection spatial angles. We consider the realistic spatial structure of trapped ion strings in a single harmonic potential formed along the axial direction of a linear Paul trap. The experimental tests are considered in the elastic scattering limit and build on the demonstrated feasibility of scalable interference from large ion strings~\cite{obs19,wol16}. The performed simulations predict that within the typical motional regimes and corresponding position uncertainties of ions, there is no significant disadvantage from the non-equidistant crystal structure compared to idealized regularly spaced chains of atomic scatterers. Following the experimental implementation limited by finite thermal position uncertainties of $^{40}$Ca$^+$~ions, we analyze the prospects of further enhancements by employing $^{138}$Ba$^+$, which provides a more suitable configuration of longer scattering wavelength and higher spatial localization at comparable trapping frequencies and the Doppler cooling limit. In addition, an analogous collection approach for spontaneously emitted photons from ion strings prepared in an entangled state of electronic excitations can be envisaged, following the methodology demonstrated previously with two-ion crystals~\cite{araneda2018interference}.

\begin{figure*}[t!]
\begin{center}
\includegraphics[width=1.5\columnwidth]{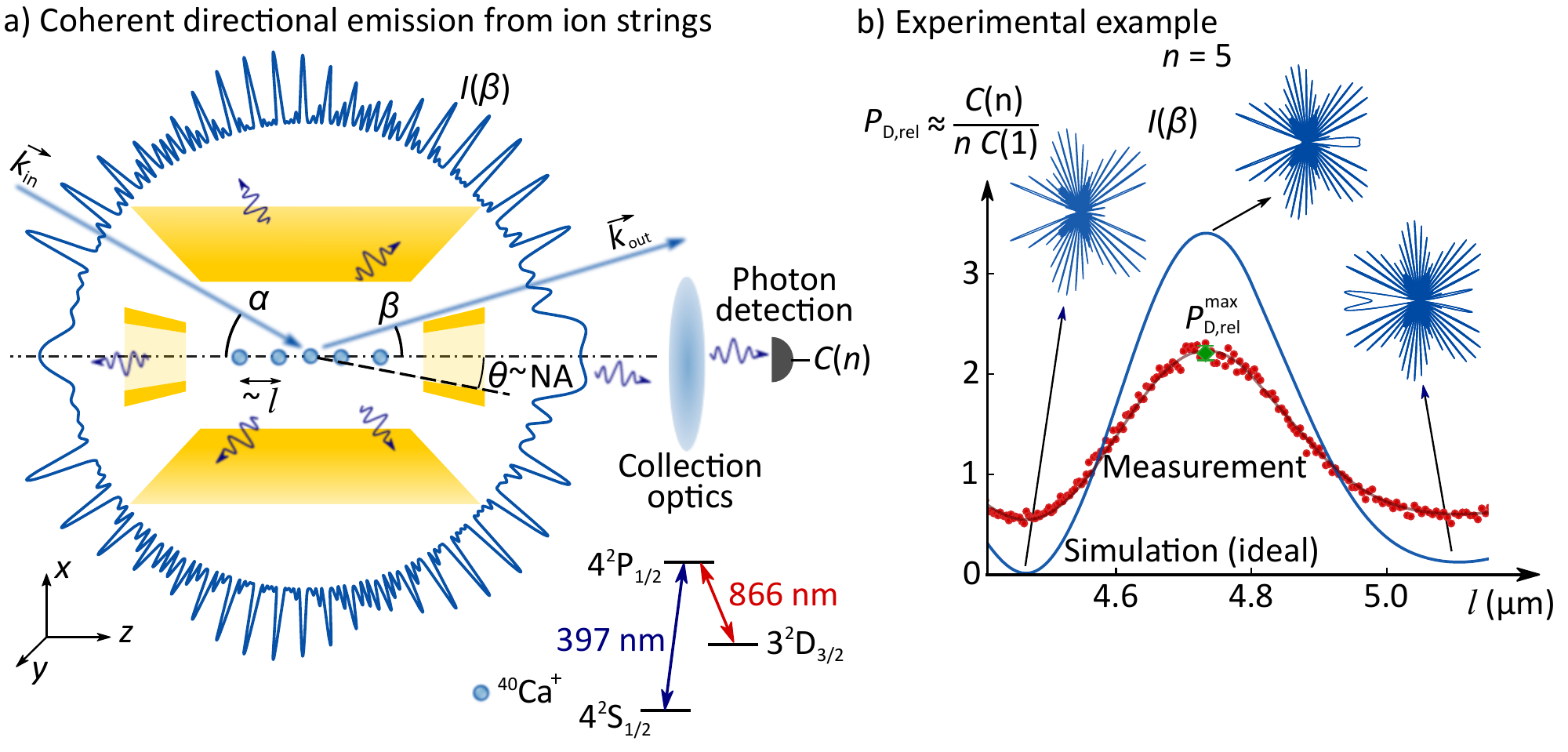}
\caption{Illustration of the coherent enhancement of collection of light from linear ion crystals in a single harmonic trapping potential. a) depicts the principle of employment of far-field angular intensity distributions $I(\beta)$ of light scattered from the excitation beam with a wave vector $\vec{k}_{\rm in}$ for maximizing the gain in the collection efficiency $P_{\rm D, rel}$ in available axial optical collection numerical aperture ${\rm NA}=\sin \theta$ along the ion crystal axis. The optimization over the spatial length scale~$l$ of the ion crystal can be tailored by controlling the axial trapping potential, which allows for direct experimental tunability and evaluation of the enhancement. The presented measurement example in b) with $n=5$ trapped $^{40}{\rm Ca}^+$ ions illustrates the feasible modification of relative detection efficiency for excitation angle $\alpha=45^{\circ}$. The corresponding simulated normalized scattering spatial patterns $I(\beta)$ depict several notable cases, including destructive emission along both axial directions, constructive in the direction of the detector, and constructive in the opposite direction, from left to right, respectively. The normalization of the detected photon rate $C(n)$ to the rate from the same number of ions in a fully incoherent scattering regime $n C(n=1)$ allows for a direct evaluation of the relative collection enhancement $P_{\rm D, rel}$.
}
\label{fig:Schema}
\end{center}
\end{figure*}

\section{Methods}

We illustrate the capability of the proposed scheme on a generic experimental scenario with broadly feasible ion trapping parameters. The scattering wavelength and excitation angle are selected according to the presented experimental tests with $^{40}$Ca$^+$, however, the results can be easily adapted to any other combination of atomic ion species and excitation geometries. The experimental demonstration employs a 3D~linear Paul trap and collection of the 397~nm photons. This corresponds to several recent experiments on coherent elastic scattering from ion strings~\cite{obs19,wol16,verde2024spin}. The trapping of linear ion strings is achieved by a radio-frequency~(RF) voltage $U_{\rm rf}\cos(\omega_{\rm rf} t)$ applied to radial electrodes and a static voltage $U_{\rm tip}$ at two axial cone-shaped hollow tip electrodes, which together provide a three-dimensional trapping potential. The frequency of the secular motion of ions along the axial direction is given by $\omega_{z} = \sqrt{2 q U_{\rm tip}\kappa/(m z^2_0)}$, where $m$ is the ion mass, $q$ is the ion charge, $2 z_0$ is the distance between the two tip electrodes, $U_{\rm tip}$ is the voltage applied to tip electrodes, and $\kappa$ is the geometrical factor. The radial motional frequencies $\omega_{r} =q U_{\rm rf}/(mr_0^2\omega_{\rm rf}\sqrt2)$ depend on the amplitude of the RF potential~$U_{\rm rf}$ and its angular frequency~$\omega_{\rm rf}$. Fig.~\ref{fig:Schema}-a) illustrates the excitation and detection geometry, which is similar to that previously employed both in neutral atom and trapped ion interference experiments~\cite{tam20,obs19}. The scattering of the 397~nm field with a red-detuned frequency~$\omega_l$ from the $4^2{\rm S}_{1/2} \leftrightarrow 4^2{\rm P}_{1/2}$ electric dipole transitions provides a simultaneous Doppler cooling of ions. The laser scatters off the ion string at an angle $\alpha\approx 45\,^\circ$ with respect to the axial trapping direction in the $z-x$ plane. The 866~nm laser beam with the same direction is used for reshuffling the population of the metastable $3^2{\rm D}_{3/2}$ manifold back to the cooling transition. The degeneracy of Zeeman states is lifted by applying a static magnetic field with a magnitude $|\vec{B}|=3.3$~Gauss along the $y$-direction. Polarizations of both excitation lasers are considered to be linear with the vector perpendicular to the applied magnetic field.

The scattered photons are collected within a spatial angle parameterized by the numerical aperture in a vacuum ${\rm NA} = \sin \theta$, where $\theta$ is half of the aperture angle.
The relevance of the axial optical access along the $z$-direction emerges from the availability of the smallest angular gradient of the intensity patterns generated by scattering light from linear ion strings along their symmetry axis.
The maximization of the coupling in the particular solid angle fraction centered along the axial direction of the trap using far-field interference critically depends on the collective constructive contribution to the coupled mode given by the relative positions of ions in the same linear harmonic potential.
The equilibrium position of $i$-th ion in the laser-cooled string can be found by minimizing the potential energy of the whole $n$-ion crystal. Since the spatial configuration in a common harmonic potential is effectively determined by a single parameter, ions positions can be conveniently parameterized using a common spatial length scale $l = (q^2/(4\pi \epsilon_0 m \omega_z^2))^{(1/3)}$, where $\epsilon_0$ is permittivity of free space, such that $z_i = l v_i$, and $v_i$ are dimensionless equilibrium positions~\cite{jam98}.
The upper limit on the axial trapping frequency is ultimately given by the requirement on the geometrically linear configuration of ions in the trap, which sets the ratio $A = (\omega_{z}/\omega_{r})^{2}$ to be lower than $A_{\rm crit}= c n^B$, where parameters $c = 2.94$ and $B = -1.8$~\cite{enz00,fis08,sch93}. However, we note that in the experiment, other factors affecting the stability of the linear crystal and the feasibility of the optimal laser cooling will further restrict the convenient range of axial motional frequencies.
The opposite - maximal spatial length scale is set to the frequency of the secular motion in the axial direction $\omega_{z}^{\rm min}= (\frac{\lambda}{4})^{-2} \frac{\hbar}{2m}$, such that the corresponding position uncertainty at the Doppler cooling limit effectively still allows for resolving optical interference from two scatterers.



We access the feasible photon collection enhancements by evaluating the far-field spatial angular dependence of optical intensity resulting from interference of light scattered elastically from a linear ion string. We consider a scattering of the beam with a plane wavefront and direction given by the excitation angle $\alpha$. The evaluation of an intensity pattern on a spherical screen represents a generic scenario, which can be applied to comparing the achievable relative collection enhancements irrespective of the particular collection optics and photon detection system. Any linear optical imaging with a corresponding input numerical aperture will merely affect the resulting spatial intensity patterns at the detector.
The complex amplitude of light field scattered from the first ion in the far-field can be written as ${U}_1 = {\varepsilon_0} e^{{\rm -i}(k r -\varphi_0)}$, where $\varepsilon_0$ is the amplitude, $k$ is the wave number, $r$ is the distance travelled by the wave from the first ion to the screen, and $\varphi_0$ is the phase of the light at the position of the first ion. Here, we omit the residual $\beta$-dependent phase offset, as the following evaluation of interference patterns depends only on the relative phase delay of the fields scattered from different ions.
The complex amplitude of light scattered from the $j$-th ion can then be expressed as $U_j =  \varepsilon_j e^{{\rm -i} (k r -\varphi_0)} e^{{\rm -i} k \Delta_d^{j,1}}$, where the path difference $\Delta_d ^{j,1} = l(v_j-v_1)(\cos{\alpha}-\cos{ \beta})$ is taken with respect to the first ion.
The far-field amplitude resulting from the scattering of the monochromatic light with a plane wavefront and a wavenumber $k$ from a string of $n$~identical two-level scatterers is given by the sum of complex amplitudes from different ions $U_{\rm tot} = \sum_{j=1}^{n} U_j$. For simplification, we assume identical amplitudes $\varepsilon_j = 1$, which is an experimentally plausible approximation in the considered axial crystal spatial lengths and feasible excitation beam widths~\cite{mielec2018atom,obs19}. The spatial intensity pattern $I(\beta) = |{U_{\rm tot}}(\beta)|^{2}$ can then be evaluated as
\begin{eqnarray}\label{eq:I_cos}
I(\beta) = |\sum_{j=1}^{n} U_j(\beta)|^2 &= |\sum_{a,b=1}^{n} U_{a}(\beta) U_{b}(\beta)|= \\   \nonumber
&=  \sum_{a,b=1}^{n} \cos(k \Delta_d^{a,b}),
\end{eqnarray}
where $\Delta_d^{a,b}$ is a path difference between fields scattered from the $a$-th ion and the $b$-th ion proportional to their mutual distance $l(v_a-v_b)$.
We note that this simplified model neglects polarization effects and corresponding orientations of atomic dipoles, as they provide on average the same enhancement for considered scattering from multi-ion crystals and to the reference average emission from a single emitter.

The optimization of collection enhancements can be quantified by the evaluation of the maximal photon flux $\Phi_{\rm NA}(n) = \int_{\Omega} I{\rm } {\rm d} \Omega$ within a given solid angle $\Omega=2\pi(1-\cos \theta)$ over experimentally feasible spatial length scales $l$. Equivalently, parametrization by the corresponding numerical aperture $\theta = \arcsin (\rm NA)$ can be employed.
Due to the circular symmetry of the scattering patterns around the $z$-axis, the photon flux can be in a paraxial limit evaluated in spherical coordinates as $\Phi_{\rm NA}(n) = \int_{\phi=0}^{2 \pi}  \int_{\theta'=0}^{\theta} I \sin \theta'  {\rm d}\theta' {\rm d}\phi =  2\pi \int_{0}^{\theta}I \sin \theta'  {\rm d}\theta'$.
Using the overall flux from an $n$~ion string $\Phi_{\rm 4 \pi}(n) = n 4\pi$,
the collection efficiency can be defined as $P_{\rm D}=\frac{\Phi_{\rm NA}}{\Phi_{\rm 4 \pi }}$, where $\Phi_{\rm NA}$ denotes the conventional parametrization by the numerical aperture~NA.
For accessing a more direct estimation of the enhancement, the relative enhancement factor $P_{\rm D, rel}$ can be defined as the collection efficiency for the $n$~ion string normalized by the collection efficiency for a single ion
\begin{equation} \label{eq:P_rel}
 P_{\rm D,rel}(n) =  \dfrac{P_{\rm D}(n)}{P_{\rm D}(n=1)} = \frac{\Phi_{\rm NA}(n)}{\Phi_{\rm NA}( n=1) n}.
\end{equation}


%
%

\section{Results} \label{sec:Results}

{\em Simulations.---} The enhancement of photon collection efficiency by the presented methodology depends on the feasibility of geometrical arrangements providing a close to constructive interference of the scattered light from many ions within the selected solid angle. The practical limits on such configurations in ion traps are imposed by the stability of linear ion crystals.
We present the simulations of feasible enhancements for up to $n = 10$~ions considering experimentally broadly feasible radial secular trapping frequencies $\omega_{r}= (2\pi) \times 5$~MHz for evaluation of the upper limit on $\omega_{z}^{\rm min}$. For higher ion numbers, insufficient axial compression of the string in the harmonic trap prevents further enhancement in the practically useful limit of NA~$>0.05$. In addition, the finite thermal motion is expected to significantly reduce the practical applicability of the employment of high ion numbers, which require increasingly lower axial trapping frequencies necessary for achieving close to a quasi-periodic structure. The excitation angle $\alpha = 45^{\circ}$ is available in many experimental setups, as this configuration allows for simultaneous efficient cooling of the axial and radial motion of ions in a linear Paul trap. However, smaller excitation angles would further decrease the sensitivity to thermal motion, and the optimal setting would correspond to an angle larger than the maximal collection angle $\theta$, such that it would closely avoid the contribution of the excitation laser light to the detected photon signal. The presented simulation constraints result in evaluated values of minimal spatial length scales ranging from $l_{\rm min} = 1.61\,\mu$m, to $l_{\rm min} = 4.23\,\mu$m for $n = 2$ and 10~ions, respectively. The value of the maximal spatial length scale $l_{\rm max} = 81.18 \,\mu$m corresponding to $\omega_{z}^{\rm min}$ is kept the same for all simulations. The photon collection enhancements have been numerically optimized over the spatial parameter~$l$.

\begin{figure}[t!]
\begin{center}
\includegraphics[width=0.8\columnwidth]{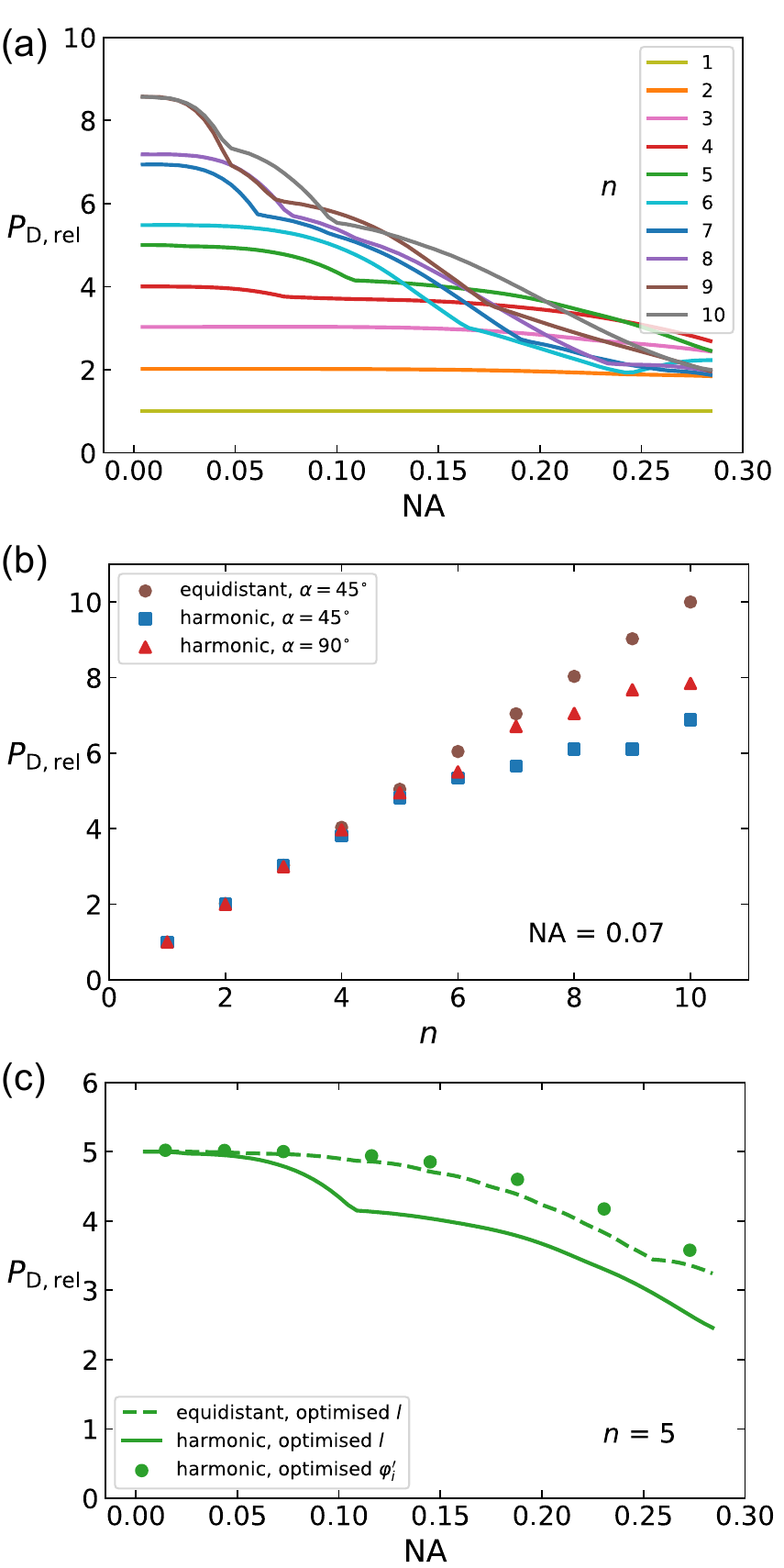}
\caption{
Simulations of the relative enhancement of the collection efficiency $P_{\rm D, rel}$ of light from linear ion crystals in a single harmonic trapping potential.
a)~shows the optimized $P_{\rm D, rel}$ for different numbers of $^{40}$Ca$^+$ ions within the practically relevant range of numerical apertures~NA of the collection optical mode. The scattering wavelength $\lambda=397$~nm and $\alpha=45^{\circ}$.
b)~depicts feasible $P_{\rm D, rel}$ for different numbers of ions and the numerical aperture ${\rm NA}=0.07$ corresponding to the employed experimental test.
Results of optimization in the harmonic potential for the two angles of incidence $\alpha = 45^{\circ}$, $\alpha = 90^{\circ}$, and for the case with regular ion spacing and $\alpha = 45^{\circ}$ (circles) are shown as full squares, triangles, and circles, respectively.
c)~illustrates the example of different optimization protocols for the string of $n=5$ ions. The optimization solely over the single trapping parameter - the spatial length scale~$l$ in the harmonic trapping potential is shown as a solid curve. The addressable tunability of individual scattering phases in the single harmonic potential allows for maximization of $P_{\rm D, rel}$ for the lowest feasible mutual distance of ions given by $l = l_{\rm min}$, with resulting enhancements shown as green points. The dashed line depicts the example of optimization for equidistant scatterers.
}
\label{fig:Calculation}
\end{center}
\end{figure}

Fig.~\ref{fig:Calculation}-a) illustrates the feasibility of significant gains in a broad range of numerical apertures. In the limit of NA~$\ll 0.1$ and small ion numbers, the enhancement $P_{\rm D, rel}$ becomes close to linear with~$n$. This is a consequence of approaching the scaling of ideal constructive interference from a quasi-periodic array of point scatterers, where $I_{\rm max}\propto n^2$, as the angular interference pattern does not approach the first destructive node within the corresponding analyzed small spatial angles. The divergence from this idealized scaling for large~$n$ is given only by residual deviations from the quasi-periodic ion positions within the available range of axial trapping frequencies. For higher numerical apertures NA~$\gg 0.1$, the advantage due to the coherent contributions gradually decreases. The phase of the interference patterns oscillates more rapidly for large scattering angles and the overall gain vanishes due to the averaging of constructive and destructive interference contributions.

As becomes emergent from the simulations in Fig.~\ref{fig:Calculation}-a), the considered maximal numerical aperture NA~$=0.3$ approaches field distributions, which include scattering observation angles~$\beta$, where closely spaced interference maxima and minima tend to average out and suppress further significant enhancements. For high ion numbers and corresponding lengths of the crystal, the angular width of the central constructive interference lobe decreases. For $n > 3$, the dependence of $P_{\rm D, rel}$ on NA displays kinks, which correspond to points of transitions between different optimal maximal constructive lobes, which result in maximal $P_{\rm D, rel}$ at different numerical apertures~NA. Generally, lower~NA allows for the employment of a narrower angular distribution of the contributing constructive lobe of the interference pattern.

The ideal quasi-periodic ion positions cannot be guaranteed on the axis for the case of four or more ions in a single harmonic trapping potential. The optimized numerical solutions presented in Fig.~\ref{fig:Calculation}-a) include many configurations, where mutual ion positions provide significant collection enhancements in a given solid angle despite residual deviations from the fully constructive interference configurations. We quantify the corresponding inefficiency of relative enhancement by analysis of the case with equal mutual distances between neighboring ions. To allow for a physically relevant comparison, the range of the mutual distances was in this case limited to the average distance of the same number of ions in the harmonic trap given by the presented limits on their motional frequencies $\omega_{z}^{\rm min}$ and $\omega_{z}^{\rm max}$. 
The example of simulation in~Fig.~\ref{fig:Calculation}-b) 
for different ion numbers and NA~$\sim 0.07$ and $\alpha = 45^{\circ}$ corresponding to our experimental setting suggests, that the ideal equidistant spacing does not provide further significant enhancement in this limit. 
The additional relative enhancement of collection efficiency achievable by employing an equidistant ion string can be quantified as $P_{\rm D, rel}^{\rm eq}/P_{\rm D, rel}^{\rm harm}=1.48$ with 9 ions in an ideal periodic lattice.
In comparison, the feasible improvement by implementation of excitation at $\alpha = 90^{\circ}$, while keeping the harmonic potential, reaches $P_{\rm D, rel}^{90^{\circ}}/P_{\rm D, rel}^{45^{\circ}} = 1.26$ for the same 9~ion string. We note that such perpendicular excitation can be additionally advantageous due to the reflection symmetry of the interference patterns, where simultaneous enhancement of coupling to both opposite axial directions can be achieved.

The particularly high spatial localization, long-term stability, and large mutual distances between ions in Paul traps allow for efficient optical addressing schemes~\cite{naegerl1999laser,debnath2016demonstration}.
Excitation of the ion string with an array of tightly focused laser beams with individually controllable~$n-1$ relative phases allows for maximization of the coupling of scattered light in the given solid angle. Alternatively, such a scenario could also be realized in the inelastic scattering regime by implementing local corrections to phases of light scattered by different ions to the collectively formed far-field interference pattern using the addressable AC-stark shifts on a string of ions prepared in a collective single-excitation Dicke states of internal electronic levels~\cite{araneda2018interference,richter2023collective}.
An example of a simulation of the optimal phase differences for the given excitation geometry and for the case of $n=5$ ions in the single harmonic potential considers $n-1=4$ independently optimized phases. Eq.~(\ref{eq:I_cos}) can then be modified to
\begin{equation} \label{eq:I_faze}
I(\beta) \sim \sum_{a,b=1}^{n} \cos[\Delta_\varphi^{a,b}+(\varphi^{\prime}_{a} - \varphi^{\prime}_{b}) ], \\
\end{equation}
where $\varphi^{\prime}_{a}$ and $\varphi^{\prime}_{b}$ are phase shifts for different ions with respect to the initial phase of the first ion in the string. Thus, $\varphi^{\prime}_{1}=0$ and $\varphi^{\prime}_{2} - \varphi^{\prime}_{n} $ are varying from~0 to~2$\pi$.
Their independent optimization in principle allows for employing the shortest feasible value of the spatial length scales $l\rightarrow l_{\rm min}$ for the given radial confinement. 
Fig.~\ref{fig:Calculation}-c) presents an example comparison of the three coupling methods for a linear string of five ions. The green data points depict the results of the approach with numerically optimized independent phase factors in scattering from different contributing ions in a single harmonic trap. The result depicted as the solid line corresponds to the equal initial phase factors for all ions and optimization over a single parameter - the spatial length scale~$l$ and is identical to the result presented in Fig.~\ref{fig:Calculation}-a). The dashed line represents the optimization over the range of distances between ions in a periodic ion string. In the limit of low numerical apertures NA~$<0.1$, the three approaches provide nearly identical collection enhancements and the harmonic trapping potential does not present any limitation. For higher solid angles, the possibility of optimization of initial phases allows for significant improvement of enhancements even beyond the equidistant case, which is due to the optimal constructive interference over given solid angles achievable at the smallest distance allowed by the trap~$l_{\rm min}$.



{\em Experimental test.---} The practical feasibility of enhancement of collection efficiency has been tested for a crystal of $^{40}$Ca$^+$ ions trapped in the linear Paul trap, as illustrated in the example presented in Fig.~\ref{fig:Schema}-a). The frequency of the RF voltage applied to radial electrodes was $\omega_{\rm{rf} } = (2\pi) \times29.9$ MHz.
The two radial motional modes had close to degenerate secular frequencies of about $(2\pi)\times 2.2$~MHz.
The axial secular frequencies $\omega_{z}$ were experimentally optimized by scanning the applied static voltage $U_{\rm tip}$ to confirm the predicted collection enhancements within ranges $\omega_z \approx (2\pi) \times (0.60 - 1.22)$~MHz for 2~ions to $\omega_z \approx (2\pi) \times (0.30 - 0.77)$~MHz for 9~ion crystal. They correspond to $l$ parameter ranging from~3.9 to 6.2$~\mu$m and from~5.3 to 9.6$~\mu$m for 2 and 9 ions, respectively.
The linear structure of the string was monitored by the camera in the radial trapping direction during the whole experiment.

The far field collection of photons scattered by ion strings in the axial direction has been based on a single plan-convex lens with a focal length of 100~mm. The limit on the effective observation spatial angle was set by the aperture of the axially positioned trapping electrode, which has been estimated from the trap design to NA~$\approx 0.07$. The collected light passed a polarization beam-splitter which transmitted the linear polarization oriented perpendicular to the direction of the applied magnetic field, such that it maximizes the portion of elastically scattered field from the $\sigma$-transitions in the detected signal.
The precise measurements of the count rate from a single ion and the background count rate were implemented to allow for the exact evaluation of the relative enhancements of coupling of photons from ions. A count rate from a single trapped ion for the laser excitation settings allowing for the stable continuous measurement and optimal Doppler cooling for all ion crystals was estimated to $C(n=1)=270 \pm 17$~counts/s.
It corresponds to a saturation parameter of the scattered 397~nm laser of about $s_{397}\approx 0.65$ and detuning $\Delta_{397} = -21$~MHz, optimized for the maximal crystal stability and visibility of the interference on the largest employed ion string with $n=9$. To allow for the reliable experimental estimation of scaling with the number of ions, the laser parameters and radial trapping frequencies were kept identical in all measurements. The excitation laser frequencies were stabilized to an optical frequency comb which provides a long-term stability~\cite{levsundak2020optical}. The measured background of $C_{\rm bg}=24\pm 5$~counts/s originated dominantly from the residual scattering of the 397~nm laser off the radial trapping electrodes.



An example of an experimental test in Fig.~\ref{fig:Schema}-b) illustrates the tunability of the collection enhancement on the scattering from $n=5$~ions. The optimization over axial potential strength was performed for all presented experimental values within the feasible ranges of the spatial length scales~$l$.
The normalized enhancement of the photon collection efficiency was then evaluated from the measured photon rates as
\begin{equation}
P_{\rm D,rel}^{{\rm exp}}(n) = \frac{C(n)-C_{\rm bg}}{(C(n=1)-C_{\rm bg})n}.
\end{equation}
The measured data were fitted using the model $P_{\rm D, rel}^{{\rm cal}}$ considering the ideal interference of the coherently scattered part according to the Eq.~(\ref{eq:I_cos}) with the probability $f_{\rm coh}$ and the complementary incoherent part of the scattered light $f_{\rm incoh}=1-f_{\rm coh}$,
\begin{equation}
P_{\rm D,rel}^{\rm exp}= f_{\rm incoh}+f_{\rm coh}P_{\rm D, rel}^{{\rm cal}}.
\label{fit_eq}
\end{equation}
The parameter $ f_{\rm coh} $ was evaluated from the local fits of the model~(\ref{fit_eq}) to the measured data to suppress the effect of measurement noise and the residual deviations of the simplified theoretical model from the experimental reality.
The resulting local maximum is depicted as a green diamond and the error bar depicts a statistically evaluated single standard deviation.

\begin{figure}[t!]
\begin{center}
\includegraphics[width=0.9\columnwidth]{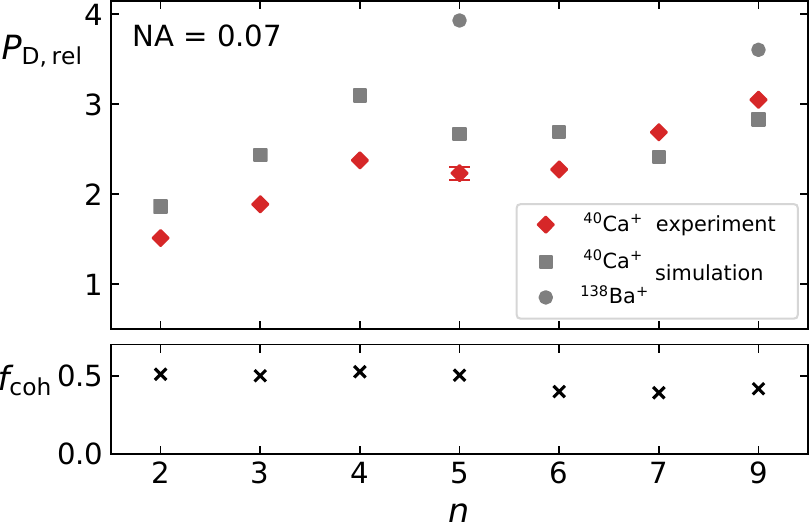}
\caption{Summary of the measured enhancements of collection efficiency for a linear chain of $^{40}$Ca$^+$ ions for the NA$\approx 0.07$. Values evaluated from the measurements shown as red diamonds can be compared with simulations shown as squares, which consider the thermal motion of ions at the Doppler cooling limit. 
The coherent fraction $f_{\rm coh}$ shown in the bottom graph was evaluated from the measured interference dependencies on the spatial length scale~$l$ using the Eq.~(\ref{fit_eq}). The error bars correspond to a single standard deviation and are smaller than the displayed data point symbols where not shown. The examples of predictions for $^{138} {\rm Ba}^{+}$ illustrate the expected enhancements for the equivalent excitation configuration.}
\label{fig:Experiment}
\end{center}
\end{figure}

Fig.~\ref{fig:Experiment} presents the summary of the measured maximal values of $P_{\rm D,rel}^{\rm exp}$ (red diamonds) and the estimated fractions of coherently scattered light (black crosses).
The values of coherent fraction $f_{\rm coh}$ estimated from the fit using the Eq.~(\ref{fit_eq}) are similar across the whole range of ion numbers, which supports the relevance of the observed scaling of $P_{\rm D, rel}$ with $n$. We note that its relative decrease by about 20~\% of the initial value for $n>5$ ions can be mostly attributed to increased spatial position uncertainties, as the optimal working points correspond to larger spatial length scales~$l$ for these number of ions. The raw collection enhancements have been measured from about $1.51\pm 0.01$ for two ions and monotonously increased to $3.05\pm 0.09$ for a nine-ion crystal.

The measured two ion enhancement value is close to what has been presented previously~\cite{araneda2018interference}. Despite working here in a much smaller NA limit, these values can be directly compared because $P_{\rm D,rel}$ is independent of~NA for a very small number of ions in the corresponding effective NA regimes, as illustrated in Fig.~\ref{fig:Calculation}-a). The intrinsic regularity of ion spacing for two and three-ion strings enhances their robustness to various geometrical factors and enables comparison among different experimental regimes and ion trapping platforms. The residual differences can be attributed mostly to thermal motional effects specific to different atomic species. The unique feasibility of testing with a higher number of ions presented in this work allows for the first observation of the scaling of experimental gains for $n>2$ trapped ions.
We note that the absolute overall detection efficiency of photon scattered from a single trapped ion estimated to $P_{\rm D,abs}= 1.7\times 10^{-2}$~\% remains relatively small mostly due to the employment of trap with the axial access corresponding to the optical solid angle of $\Omega \approx 0.015$ and the finite detection efficiency of employed single-photon counting module of $\eta_{\rm SPCM}\approx 50$~\%. However, the relative enhancement for 9~ions gives final $P_{\rm D,abs} = 0.051 \pm  0.001$~\%,
which is close to the optimized value of the overall detection efficiency in radial direction $P_{\rm D,abs} = 0.06$~\% measured with the collection objective with much higher NA of $\approx 0.3$ covering about $2$~\% of the full solid angle in the same apparatus and in the analogous photon detection settings~\cite{obvsil2018nonclassical}. These comparisons provide an experimental validation of the applicability of the proposed approach for diverse applications requiring efficient photon collection from ions, including implementation of particularly challenging experiments requiring multi-photon detection events from many ions~\cite{araneda2018interference,singh2024coherent}.

The finite saturation parameter and the multilevel electronic level structure of $^{40}{\rm Ca}^{+}$ reduce the observed enhancements, which amounts to the residual difference between the measured and simulated data points. The contribution of the inelastically scattered light has been estimated to be about 38~\% from a dark resonance spectroscopy. In addition, the steady-state population of the metastable $3^2{\rm D}_{3/2}$ manifold of about 13~\% effectively corresponds to a random switching of the contribution of different ions to the interfering signal. However, we remind that these effects are intrinsic to the experimental tests relying on the elastic scattering and should not affect the collection efficiency of photons scattered from the ion crystal prepared in entangled $|W\rangle$ states with coherently shared electronic excitation~\cite{araneda2018interference,richter2023collective,cole2021dissipative,haffner2005scalable}.

The reduction of the interference visibility due to the position uncertainty of ions has been estimated to be the most significant deterioration process, similar to previous demonstrations of interference from strings of trapped ions~\cite{ita98,araneda2018interference,wol16,obs19}.
The impact of motion increases for low trapping frequencies and corresponding larger spatial length scales~$l$ due to the greater mean position uncertainty of ions, which also becomes emergent from the measured maximal enhancements.
Considering the position uncertainties of thermal motional states, the equation~(\ref{eq:I_cos}) can be modified to
\begin{equation}
\label{eq:I_motion}
I \sim \sum_{a,b=1}^{n} e^{-\frac{1}{2}k_{\rm eff}^2\sigma_{a,b}^2}
\cos \Delta_{\varphi}^{a,b}.
\end{equation}
Here $\vec{k}_{\rm eff} = \vec{k}_{\rm out} - \vec{k}_{\rm in}$, $\vec{k}_{\rm out}$ and $\vec{k}_{\rm in}$ are the scattered and incident wave vectors, respectively,
and $\sigma_{a,b}$ is the standard deviation of the mutual distance between $a$-th and $b$-th ions considering full decomposition to $n$~normal axial motional modes. The corresponding simulations of collection enhancements affected by motional dephasing are shown in the Fig.~\ref{fig:Experiment} as grey squares. The average reduction of the relative enhancement for $n = 2$ to~6 ion crystals due to the thermal motion at the Doppler cooling limit has been evaluated to around 25~\%, while the estimated suppression becomes clearly more significant with the increasing spatial length scale for higher ion numbers.
The enhancement for the given trapping parameters is thus expected to be further significantly improved when employing heavier ion species, as the variance of the mutual ion positions is given by $\sigma_{a,b}^2 \sim \frac{\hbar}{2m\omega_{z}}$, or by employing higher scattering wavelength. 
Fig.~\ref{fig:Experiment} includes examples of simulations with $^{138} {\rm Ba}^{+}$ considering scattering at the $4^2{\rm S}_{1/2}\leftrightarrow 4^2{\rm P}_{1/2}$ dipole transition with $\lambda\approx 493$~nm, shown as grey circles.
The two presented data points represent notable cases of the largest feasible relative enhancement in comparison with the measurements on $^{40}{\rm Ca}^{+}$ for $n=5$ and with the highest measured $P_{\rm D, rel}^{\rm exp}$ for $n=9$.
For $n=5$, the simulation predicts additional improvement $P_{\rm D, rel}^{\rm Ba}/P_{\rm D, rel}^{\rm Ca}\approx 1.45$, i.e. a total enhancement of about $P_{\rm D, rel}^{{\rm Ba}} \approx 3.93$.
The predicted optimum for $n=5$ is the consequence of the enhanced sensitivity to motion for larger $n$, which require longer spatial length scales to approach quasi-periodic crystal structure. At the same time, these longer crystals result in spatial interference patterns with a much smaller angular width of the first constructive lobe. These effects jointly contribute to the decreasing of maximal gain $P_{\rm D,rel} ^{\rm exp}/n$ for high~$n$.

\section{Conclusions}
\label{sec:Conclusion}

The presented scheme for enhancement of collection efficiency from linear ion strings employs scattering geometry that is intrinsic to diverse modern linear ion traps. Its simulations confirm the feasibility of the close to an ideal linear gain by a factor of~$n$ in the collection efficiency of light scattered from up to $n=5$ ions for small numerical apertures NA~$<0.1$, and predict still significant enhancements even for higher ion numbers and numerical apertures, still within the experimentally feasible position uncertainties and axial motional frequencies. The scheme provides the most attractive enhancements within the range of numerical apertures~$0.05<{\rm NA}<0.2$, where the maximum gain can be in the ideal case of fully coherent scattering  approached already for relatively small ion numbers $n \approx 5$. They allow for the employment of small minimal spatial length scales~$l_{\rm min}$ for the given radial secular frequency and thus naturally provide optimal enhancement configurations due to the resulting interference patterns with smaller angular gradients. For higher ion numbers, it becomes unfeasible to achieve a sufficient axial compression of the string in a single linear harmonic trapping potential, and the corresponding minimal spatial length scales lead to the lower maximal collection enhancements. Although the optimization in the case of equidistant emitters predicts the feasibility of further enhancements, these gains can be in principle fully eliminated by control of individual scattering phases in strings of ions prepared in states possessing a collective coherent spin excitation~\cite{araneda2018interference}.

The realized experiment demonstrated enhancements ranging from $\approx 1.51$ for the two~ion crystal to $\approx 3.05$ for the nine~ion crystal, limited dominantly by the residual thermal motion, finite saturation parameters, and multilevel effects. While the detrimental impact of the inelastically scattered light and multilevel effects would be practically absent in the schemes employing the collectively shared spin excitation~\cite{araneda2018interference,richter2023collective}, the position uncertainty can be further improved by employing ions with large atomic mass, employment of transitions with longer wavelengths, further decrease of the input scattering angle, or their combination.

The presented approach enables a broadly applicable efficient photon collection from ions for small spatial angles. When compared to the free space high-NA collection setups, the considered small solid angle limit provides an inherent advantage for the reduction of wavefront aberrations of the collected light with simple paraxial collection optics, which promises perspectives in interferometric applications using collectively coupled ion crystals~\cite{duan2001long,slodivcka2013atom}. As a large focus depth of the small-NA fluorescence collection allows for simultaneous observation of light from many ions, the available control of the large trapped ion quantum registers~\cite{pogorelov2021compact,kranzl2022controlling} can allow for mapping of the internal state of any of the ions on the direction of the scattered light~\cite{araneda2018interference,richter2023collective}.

\section{Acknowledgments}
D.~B. and D.~T. are grateful for national funding from the MEYS under grant agreement No.~731473 and from the QUANTERA ERA-NET cofund in quantum technologies implemented within the European Union’s Horizon 2020 Programme (project PACE-IN, 8C20004). A.~K. acknowledges the support of the Czech Science Foundation under the project GA21-13265X. D.~T. and A.~L. were supported by national funding from the MEYS under the project CZ.02.01.01/00/22 008/0004649. O.~Č., T.~P., and L.~S. were supported by the Ministry of Interior of the Czech Republic within the Programme OPSEC under the project~VK01030193. We thank Romain Bachelard and Philipp Schindler for fruitful discussions.

\end{document}